\documentclass[letter]{aa} 

\usepackage{graphicx}
\usepackage{txfonts}
\usepackage{hyperref}
\hypersetup{colorlinks=true, urlcolor=blue, linkcolor=blue, citecolor=blue}
\usepackage{mathtools}
\usepackage{mathabx}

\begin{document} 

\title{Testing protoplanetary disc evolution with CO fluxes}
\subtitle{A proof of concept in Lupus and Upper Sco}

\author{
Francesco Zagaria\inst{1,2}
\and Stefano Facchini\inst{2,3}
\and Anna Miotello\inst{2}
\and Carlo F. Manara\inst{2}
\and Claudia Toci\inst{2}
\and Cathie J. Clarke\inst{1}
}

\institute{
Institute of Astronomy, University of Cambridge, Madingley Road, Cambridge CB3 0HA, UK \\ \email{fz258@cam.ac.uk}
\and European Southern Observatory, Karl-Schwarzschild-Strasse 2,
85748 Garching bei München, Germany
\and Dipartimento di Fisica, Universit\`a degli Studi di Milano, Via Celoria 16, I-20133 Milano, Italy
}

\date{Received ...; accepted ...}

\abstract
{
The Atacama Large Millimeter/submillimeter Array (ALMA) revolutionised our understanding of protoplanetary discs. However, the available data have not given conclusive answers yet on the underlying disc evolution mechanisms: viscosity or magnetohydrodynamic (MHD) winds. Improving upon the current results, mostly based on the analysis of disc sizes, is difficult because larger, deeper, and higher angular resolution surveys would be required, which could be prohibitive even for ALMA. In this Letter we introduce an alternative method to study disc evolution based on $^{12}$CO fluxes. Fluxes can be readily collected using less time-consuming lower resolution observations, while tracing the same disc physico-chemical processes as sizes: assuming that $^{12}$CO is optically thick, fluxes scale with the disc surface area. We developed a semi-analytical model to compute $^{12}$CO fluxes and benchmarked it against the results of \texttt{DALI} thermochemical models, recovering an agreement within a factor of three. As a proof of concept we compared our models with Lupus and Upper Sco data, taking advantage of the increased samples, by a factor 1.3 (Lupus) and 3.6 (Upper Sco), when studying fluxes instead of sizes. Models and data agree well only if CO depletion is considered. However, the uncertainties on the initial conditions limited our interpretation of the observations. Our new method can be used to design future ad hoc observational strategies to collect better data and give conclusive answers on disc evolution.
} 


\maketitle


\section{Introduction}\label{sec:introduction}
Over the last decades two disc evolution models, viscous theory \citep{Lynden-Bell&Pringle1974,Shakura&Sunyaev1973} and the magnetohydrodynamic wind (MHD-wind) scenario \citep{Blandford&Payne1982}, have been proposed \citep{Manara_ppvii}. 
According to the viscous evolution model, the disc angular momentum is conserved and redistributed by turbulence: while a small fraction of the disc mass moves to larger sizes, the bulk is accreted. Instead, in the MHD-wind scenario, powerful magnetothermal winds are launched from the disc, allowing   accretion to efficiently remove angular momentum. In addition to these mechanisms, 
thermal winds, not instrumental in driving the accretion process, are thought to play a key role in the disc dispersal phase \citep{Pascucci_ppvii}, complicating the picture. Discriminating between these two scenarios requires large surveys targeting populations of discs of different ages in order to compare models and data in a statistical sense. In recent years, the Atacama Large Millimeter/submillimeter Array (ALMA) observed several nearby star-forming regions (SFRs) \citep[e.g.][]{Ansdell2016,Ansdell2018,Pascucci2016,Barenfeld2016,Cieza2019,Cazzoletti2019} at moderate resolution (0.25 to 0.50 arcsec) and sensitivity (0.1 to 0.4 $M_\Earth$), measuring fluxes and sizes for tens of discs \citep{Manara_ppvii,Miotello_ppvii} from dust and CO rotational transitions. 

Disc sizes have been particularly useful to study disc evolution because of the different trends predicted by models: while viscous discs are expected to get larger with time, in the MHD-wind scenario discs either remain the same or shrink \citep{Manara_ppvii}. In the case of dust, \citet{Rosotti2019} predicted that the disc radius (enclosing  95\% of the total dust flux) expands with time in viscous models. However, if present, this behaviour can only be detected in very deep surveys, with a sensitivity that is  fifty times better  than in the available data. This sensitivity can be reached with roughly five hours on-source at an intermediate resolution (0.6 to 0.7 arcsec), which would be prohibitive for any future ALMA survey targeting hundreds of discs. 
\citet{Zagaria2022} extended the work of \citet{Rosotti2019}, showing that this same factor of fifty is needed to distinguish between viscous and  MHD-wind evolution. Furthermore, a direct comparison between models and data is made more difficult by the presence of substructures \citep{Toci2021,Zormpas2022,Zagaria2022} since the observed sizes may trace the effects of disc-planet interactions rather than disc evolution.

In the case of gas, following up on the early work of \citet{Najita&Bergin2018}, \citet{Trapman2020} used complex thermochemical models to show that small discs with low viscosities can explain most of the observationally inferred disc sizes in Lupus, but they spread too much to reproduce more compact discs in Upper Sco. MHD-wind models, instead, are broadly consistent with the gas disc sizes measured in both SFRs \citep{Trapman2022}. However, this comparison is affected by two main uncertainties: the small samples, particularly at the age of Upper Sco \citep{Barenfeld2017}, and the amount of carbon depletion. When the carbon abundance falls below $x_{\rm CO}\approx10^{-6}$, \citet{Trapman2022} showed that discs observed with low sensitivity could look up to 70\% smaller or be unresolved. To mitigate this problem, integration times of one hour per source would be needed, which is challenging for large surveys.

However, targeting disc sizes is not the only possible strategy to study disc evolution. Here we introduce an alternative method based on $^{12}$CO fluxes. Assuming that $^{12}$CO emission is optically thick, CO fluxes scale as the disc surface area (i.e. the radius squared), suggesting that modelling fluxes is an indirect way of studying sizes since they would trace the same physico-chemical processes in the disc. This assumption is supported by  both models \citep{Trapman2019,Trapman2020,Trapman2022,Miotello2021} and by the  data. For example, \citet{Long2022} showed that the observationally inferred CO fluxes and sizes correlate well, with $R_{\rm CO} \propto F_{\rm CO}^{0.52 \pm 0.05}$ (see also \citealt{Sanchis2021}). Observing fluxes instead of sizes is less time consuming, firstly, because one would aim to detect, but not necessarily resolve, a target, and  secondly, because there would be no need for very deep surveys targeting the faint outer disc regions that contribute marginally to the disc brightness. In this Letter we introduce a simple semi-analytical prescription to compute $^{12}$CO disc fluxes under the optically thick assumption. We benchmark this prescription against a grid of full radiative transfer simulations and show that they agree, on average, within a factor of three. Then, as a proof of concept, we compare these models with Lupus and Upper Sco data, highlighting the main limitations of the available datasets and the foreseen improvements with future dedicated surveys.

This Letter is organised as follows. In Sect.~\ref{sec:methods} we introduce our semi-analytical method. In Sect.~\ref{sec:results} we run a disc population synthesis model and compare viscous and MHD-wind predictions with Lupus and Upper Sco data. Our results are discussed in Sect.~\ref{sec:discussion}, and in Sect.~\ref{sec:conclusions} we draw our conclusions. The code developed for this work is publicly available on \href{https://github.com/fzagaria/COpops.git}{\texttt{github}}.

\section{Methods}\label{sec:methods}
Here we summarise our assumptions and final equation to compute CO fluxes (see Sect.~\ref{app:model} for the full derivation). 

We considered $^{12}$CO emission to be optically thick and in local thermodynamical equilibrium. Under these assumptions 
\begin{equation}\label{eq:2.1}
    F_{\rm CO} = \dfrac{\cos i}{d^2}\int_{R_{\rm in}}^{R_{\rm out}}\int_0^\infty I_\nu 2\pi RdRd\nu,
\end{equation}
where $R$ is the cylindrical disc radius, $i$ the disc inclination, and $d$ its distance from the observer. The brightness profile in Eq.~\ref{eq:2.1} can be written as
\begin{equation}
    I_\nu = B_{\nu_0}(T)\exp\left\{-\dfrac{m_{\rm CO}c^2(\nu-\nu_0)^2}{2k_{\rm B}T\nu_0^2}\right\}\dfrac{c}{\nu_0},
\end{equation}
where $B_{\nu_0}$ is the black-body emission at temperature $T$ and frequency $\nu_0$, and the exponential term gives the thermal broadening of the line \citep{Rybicki&Lightman1986}. Here $m_{\rm CO}$ is the $^{12}$CO molecular mass, $c$ the speed of light, and $k_{\rm B}$ the Boltzmann constant. We adopted a power-law temperature profile with exponent $-0.5$ and normalisation $87.5\,{\rm K}$ at $20\,{\rm au}$, in agreement with the inferences of \citet{Law2021,Law2022,Law2023}. Our disc inclination was fixed to the sky-averaged value of $\cos i=\pi/4$. 

We adopted $R_{\rm in} = 10^{-2}\, {\rm au}$ and $R_{\rm out}=R_{\rm CO}$, the radius where the gas surface density equals the column density, $N_{\rm CO}=5\times10^{15}\, {\rm cm}^{-2}$ \citep[a density slightly larger 
than the standard result of][]{vanDishoeck&Black1988}, where $^{12}$CO is not efficiently self-shielded against photodissociation and is quickly removed from the gas phase. To compute the gas surface density corresponding to $N_{\rm CO}$, we assumed the same carbon abundance of the diffuse ISM, $x_{\rm CO}=10^{-4}$. Although it is rather crude, this method is supported by the work of \citet{Trapman2019}, who showed that $R_{\rm CO}$ encloses all of the CO emission and is in good agreement with the results of complex thermochemical models. To test our method, we benchmarked our sizes and fluxes against the results of the thermochemical models of \citet{Miotello2016} and \citet{Trapman2020,Trapman2022}, run using the code \texttt{DALI} \citep{Bruderer2012,Bruderer2013}. The results of this exercise are extensively discussed in Sect.~\ref{app:comparison}, where we show that our face-on fluxes underestimate \texttt{DALI} ones by a factor of three.

\section{Population synthesis}\label{sec:results}
We give a proof of concept  of this new method comparing our semi-analytical predictions with the available Lupus (age $\lesssim3$ Myr, \citealt{Luhman&Esplin2020}) and Upper Sco (age 5 to 10 Myr, \citealt{Luhman2020}) data. A quick description of the datasets can be found in Sect.~\ref{app:sample}. Here we note that even with the limited data available, working with fluxes instead of sizes increases the samples by a factor of 1.3 (48 instead of 36 sources) in Lupus and  by a substantial factor of 3.6 (32 instead of 9 sources) in Upper Sco. For this comparison we relied on a disc population synthesis approach: we prescribed a set of initial conditions and evolved our models in the viscous or MHD-wind case under the assumption that these two SFRs can be regarded as subsequent evolutionary stages of the same population (i.e. they have the same initial conditions). Unfortunately, these initial conditions are either unknown or very uncertain (e.g. they were inferred neglecting any contribution of dust to disc evolution; \citealt{Lodato2017,Tabone2022b}). Future more accurate distributions will allow   more reliable comparisons between evolutionary models and data.

\begin{figure*}
    \centering
    \includegraphics[width=0.975\textwidth]{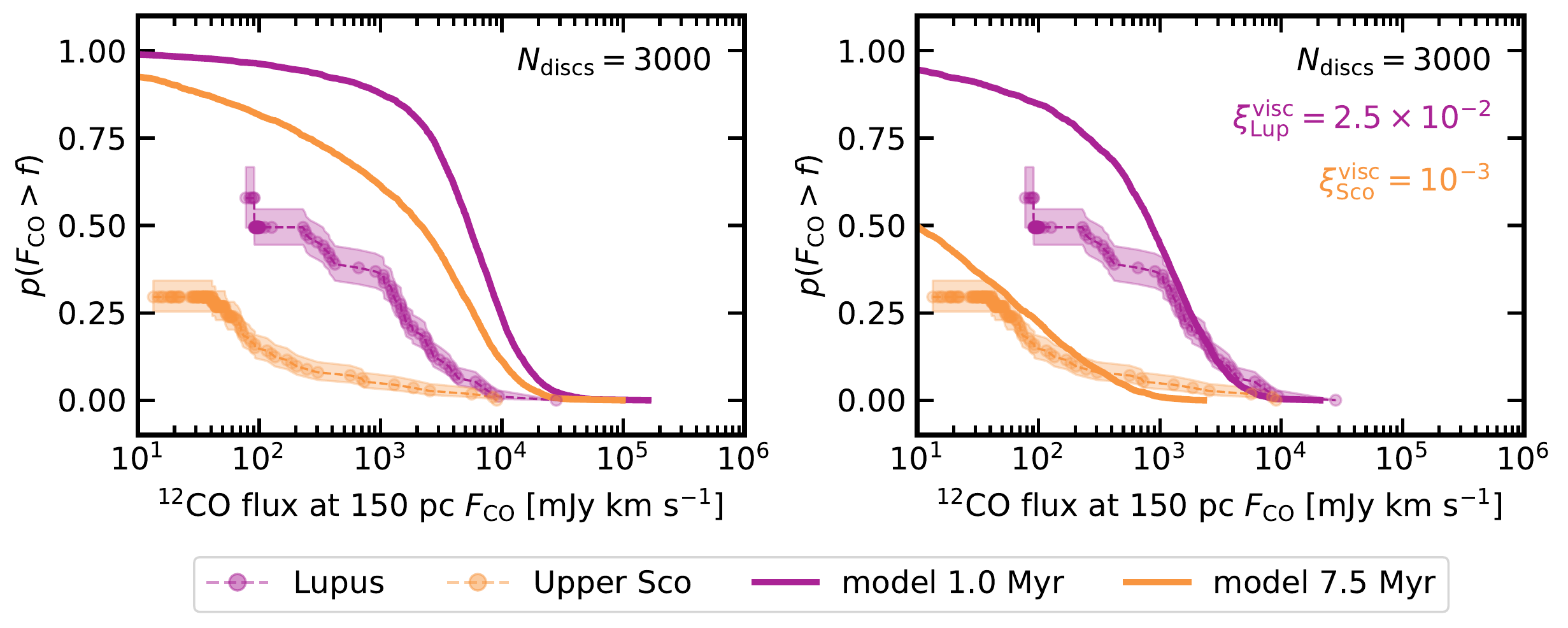}
    \caption{Comparison of the data (patches) and viscous model (solid lines) survival functions at the age of Lupus (purple) and Upper Sco (orange). \textbf{Left panel:} Standard assumptions. \textbf{Right panel:} Reduced gas column density. Fudge factors ($\xi^{\rm visc}$, top right corner) are needed to match the data.}
    \label{fig:3.1}
\end{figure*}

\subsection{Viscous case}
We used the \citet{Lynden-Bell&Pringle1974} analytical solution to compute the CO radius. In this case, the surface density at a given time is a function of the viscous timescale ($t_{\rm acc}$), initial disc mass ($M_0$), and scale radius ($R_0$). We assumed the viscous timescale to be distributed as $\log(t_{\rm acc}/{\rm yr}) = \mathcal{N}(5.8,\,1.0)$, where the notation $\mathcal{N}(\mu,\,\sigma)$ stands for a Gaussian distribution with mean $\mu$ and variance $\sigma^2$. This distribution was inferred by \citet{Lodato2017} fitting the Lupus data in the $\dot{M}_{\rm acc}-M_{\rm disc}$ plane, under the assumption that viscosity is an increasing function of the disc radius with exponent $\gamma=1.5$. For the initial disc mass distribution we considered $\log(M_0/M_\odot) = \mathcal{N}(-2.7,\,0.7)$, similarly to \citet{Lodato2017}. Even though young discs are known to be small \citep{Maury2019,Maret2020,Tobin2020}, their initial disc size distribution is not well constrained. To take into account possible envelope contributions, we adopted the best fit $R_{\rm disc}\approx R_0$ distribution of 25 Class 0 objects in Orion (VANDAM, \citealt{Tobin2020}) based on the radiative transfer models of \citet{Sheehan2022}, under the assumption that gas and dust are co-located at such young ages: $\log(R_0/{\rm au}) = \mathcal{N}(1.55,\,0.4)$.
Finally, we assumed an age of $\log(t/{\rm yr}) = \mathcal{N}(5.9,\,0.3)$ for Lupus (corresponding to our choice of $t_{\rm acc}$; see \citealt{Lodato2017}) and $7.5\,{\rm Myr}$ for Upper Sco.

We checked the $\alpha_{\rm SS}$ \citep{Shakura&Sunyaev1973} distribution associated with our initial conditions (for a $M_\odot$ star):
\begin{equation}
    \alpha_{\rm SS} = 0.67\times10^{-3}
    \left(\dfrac{R_0}{10\,{\rm au}}\right)\left(\dfrac{t_{\rm acc}}{1\,{\rm Myr}}\right)^{-1}\left(\dfrac{h_0}{0.1}\right)^{-2}.
\end{equation}
Here we considered $h$ to be a power law with exponent $0.25$ and normalisation $h_0=0.1$ at $10\,{\rm au}$, in line with the results of \citet{Zhang2021}. Our choices of $t_{\rm acc}$ and $R_0$ give a distribution of $\log\alpha_{\rm SS}\approx\mathcal{N}(-2.42,\,1.06)$.

Our results are displayed in Fig.~\ref{fig:3.1}. Measured fluxes are shown as purple and orange patches for Lupus and Upper Sco; the survival functions and their $1\sigma$ spread were computed using the Kaplan-Meier estimator for left-censored  datasets (see Sect.~\ref{app:sample}). The survival functions for $N_{\rm discs} = 3000$ models are plotted as solid lines of the same colours. Our results under standard assumptions (see Sect.~\ref{sec:methods}) are presented in the left panel. To get a better insight into these flux distributions, we follow the evolution of the median disc (i.e. the disc whose initial conditions are the median of our assumed distributions). This disc spreads viscously, getting bigger and brighter, until an inversion time $t_{\rm inv}$ (Eq.~12 of \citealt{Toci2023}). Then, the part of the disc that is viscously expanding falls below the CO photodissociation threshold, making the disc smaller and fainter. Our Upper Sco models are fainter than Lupus models because more discs (particularly those with larger $R_0$, lower $M_0$, and shorter $t_{\rm acc}$) went past their inversion time. Nevertheless, our models are $\gtrsim10$ times brighter than the data. To reconcile models and observations we introduced a column density fudge factor $\xi$, that makes photodissociation more efficient: $N_{\rm CO}\rightarrow N_{\rm CO}/\xi$. A discussion on the possible physico-chemical interpretation of such a factor can be found in Sect.~\ref{sec:discussion}. Our results are displayed in the right panel of Fig.~\ref{fig:3.1}, for $\xi^{\rm visc}_{\rm Lup}=2.5\times10^{-2}$ and $\xi^{\rm visc}_{\rm Sco}=10^{-3}$: these fudge factors are able to reconcile models and observations both at the age of Lupus and Upper Sco. However, for the faintest discs in Lupus and the brightest in Upper Sco a smaller (larger) correction factor would be required. This effect can also be due to warmer (colder) discs than our average temperature profile. We note that these fudge factors are not an artefact of our initial conditions; in other words, we found no \textit{\emph{sensible}} combination  of the initial parameters able to viscously reproduce both Lupus and Upper Sco observations with $\xi^{\rm visc}_{\rm Lup}=\xi^{\rm visc}_{\rm Sco}=1$.

\subsection{MHD-wind case}
We used the \citet{Tabone2022a} analytical solution with constant magnetic field strength ($\omega=1$) to compute the CO radius. This solution can reproduce both the disc fraction decay with time and the Lupus data in the $\dot{M}_{\rm acc}-M_{\rm disc}$ plane \citep{Tabone2022b}. In this case the surface density at a given time is a function of the accretion timescale ($t_{\rm acc}$), the initial disc mass ($M_0$), the initial scale radius ($R_0$), and the lever-arm parameter ($\lambda$). Because the wind-driven prescription accounts for disc dispersal after a finite time, knowledge of the disc fraction distribution can be used to infer a distribution of $t_{\rm acc}$ (see Eq.~A.2 of \citealt{Tabone2022b}). Following \citet{Tabone2022b}, our initial disc mass distribution is log-normal with 1.0~dex spread and centred on $M_0=2\times10^{-3}\,M_\odot$, with corresponding mass ejection-to-accretion ratio $f_{\rm M} = 0.6$. We chose the same initial radius distribution of the viscous case. Then, the lever-arm parameter distribution can be computed from $R_0$ and $f_{\rm M}$ (see \citealt{Tabone2022b}, where we fixed the innermost wind launching radius to $1\, {\rm au}$). The parameter $\lambda$ is distributed with $\mu\approx4.8$ and $\sigma\approx0.95$. Finally, we assumed an age of $2\, {\rm Myr}$ for Lupus (corresponding to our choice of $M_0$; see \citealt{Tabone2022b}) and $7.5\,{\rm Myr}$ for Upper Sco.

\begin{figure*}
    \centering
    \includegraphics[width=0.975\textwidth]{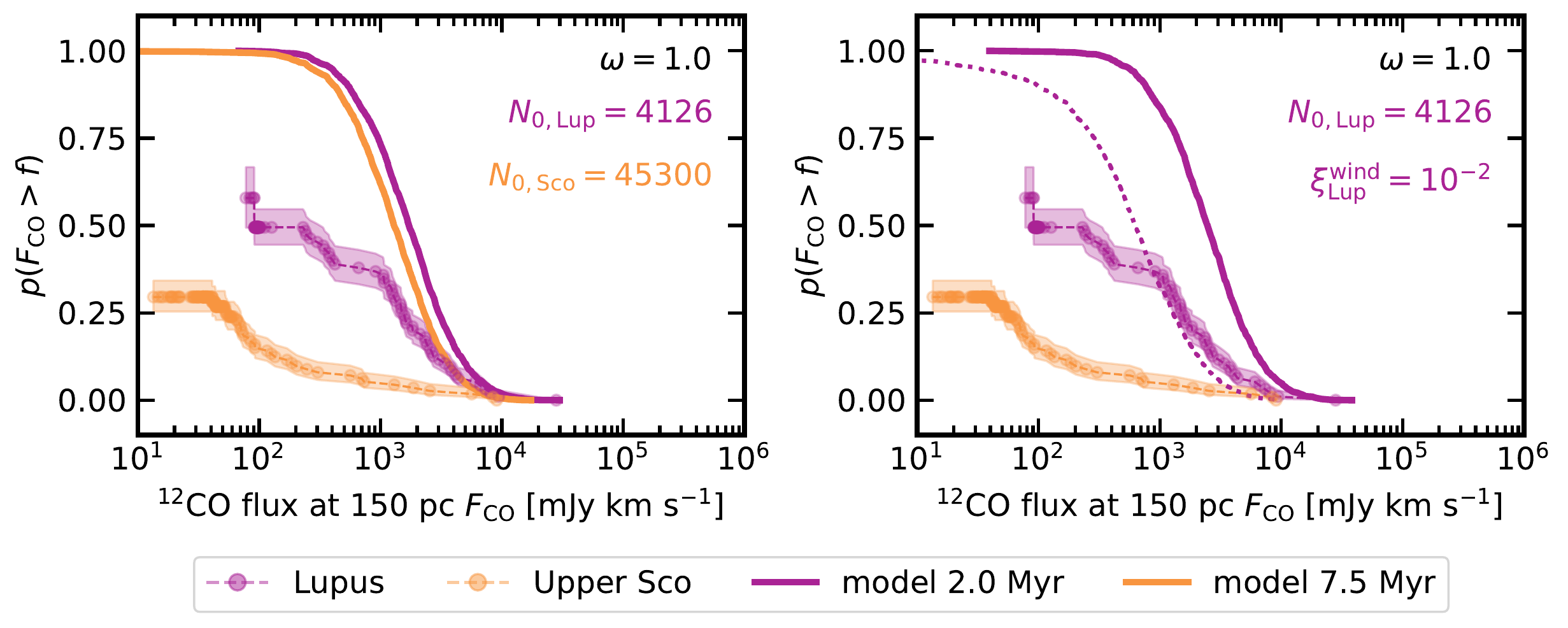}
    \caption{Comparison of the data (patches) and MHD-wind model (solid lines) survival functions at the age of Lupus (purple) and Upper Sco (orange). \textbf{Left panel:} Standard assumptions. A mass-dependent depletion factor needs to be invoked to match the data. \textbf{Right panel:} Larger initial disc size distribution in Lupus. A constant fudge factor can reproduce the data (dotted line for $\xi^{\rm wind}_{\rm Lup}=10^{-2}$).}
    \label{fig:3.2}
\end{figure*}

Our results are displayed in the left panel of Fig.~\ref{fig:3.2}, using the same symbols as Fig.~\ref{fig:3.1}. As noted above, these MHD-wind models have a finite lifetime which was chosen to reproduce the observed age dependence of the disc fraction in SFRs. Since Upper Sco is older than Lupus and their samples are of similar sizes (Sect.~\ref{app:sample}), a larger number of initial discs is needed to reproduce the number of sources observed in the former region: $N_{\rm 0,Lup} = 196$ and $N_{\rm 0,Sco} = 2490$ (in Fig.~\ref{fig:3.2} 20 times more models are shown to better explore the initial conditions). In the MHD-wind case the median disc evolves slowly and its radius is constant with time, until $t/t_{\rm depl}\gtrsim95\%$, when fast dispersal takes place and the disc gets abruptly smaller and dimmer. Consequently, we expect our CO flux distributions to be mostly dependent on the initial disc size. A clear difference with the viscous case is that the brightest MHD-wind models are almost as luminous as the brightest Lupus and Upper Sco data, while for fainter and fainter discs the discrepancy between models and data progressively increases. As a consequence, a constant fudge factor cannot reconcile models and observations; a disc mass dependent correction would need to be invoked. We obtained similar results for $\omega=0.5$ and the initial parameters explored by \citet{Tabone2022b}.

\section{Discussion}\label{sec:discussion}
In this Letter we assumed that the Lupus and Upper Sco disc populations could be considered a respectively younger and older evolutionary stage of the same initial disc population. Under this hypothesis, we ran disc population synthesis models from sensible initial conditions and compared our CO flux distributions with the data.  

To match models and data, in the viscous case we introduced a column density fudge factor that increases the CO photodissociation efficiency. This factor can be interpreted as the outcome of some processes depleting CO in protoplanetary discs \citep{Miotello_ppvii}. The low gas masses estimated from CO in Lupus and Chamaeleon~I discs \citep{Ansdell2016,Ansdell2018,Miotello2017,Long2017} indicate that protoplanetary discs are fainter than expected in CO. This is supported by the few direct measurements of disc masses based on HD rotational line transitions, that require CO to be depleted by factors between 5 and 200 \citep{Bergin2013,Favre2013,McClure2016,Schwarz2016}. Two main processes have been proposed to explain CO underabundance: (i) because of CO chemical (gas- or ice-phase) conversion or evolution, carbon would be sequestered into more complex species, like CO$_2$ or hydrocarbons, that can freeze-out onto grains at higher temperatures than CO \citep[e.g.][]{Bosman2018,Schwarz2018}; (ii) CO freeze-out on dust and subsequent grain growth into larger bodies that no longer participate in gas-phase chemistry would lock carbon up in the disc midplane and transport it radially \citep[e.g.][]{Krijt2016,Krijt2018,Powell2022}. A combination of the two processes is most likely to take place \citep[e.g.][]{Booth2017,Krijt2020} and is needed to explain depletion factors of 100 on a timescale of 1 to $3\, {\rm Myr}$ inferred from the comparison between Class I and Class II discs \citep{Zhang2020}. The carbon depletion scenario is supported by the detection of C$_2$H in several protoplanetary discs, which can be explained by carbon and oxygen depletion with ${\rm C/O}\gtrsim1$ \citep{Bergin2016,Cleeves2018,Miotello2019,Bosman2021}. This result is consistent with the carbon-to-oxygen ratio inferred from the available $N({\rm CS})/N({\rm SO})$ upper limits \citep{Semenov2018,Facchini2021,LeGal2021}.

The fudge factor we introduced in the viscous case to explain Lupus data is in line with the observationally inferred CO depletion factors of two orders of magnitude \citep{Miotello_ppvii}. Notably, in addition to CO fluxes, these carbon-depleted models can also reproduce very well the $\dot{M}_{\rm acc}-M_{\rm disc}$ correlation (by construction) and the CO size-luminosity correlation (see Sect.~\ref{app:long2022}). Here we caution that our $\xi^{\rm visc}_{\rm Lup}$ should be interpreted as a population average rather than an absolute depletion factor; other factors (e.g. a spread in the stellar mass and luminosity, dust evolution, and different source inclinations) can effect CO depletion, either by changing disc chemistry or the observed luminosities. The larger correction needed in the case of Upper Sco, instead, could be explained in three different ways. Firstly, CO is more depleted in Upper Sco than in Lupus. This hypothesis is supported by several evolutionary models, where the carbon depletion factor increases with time \citep{Krijt2020,Powell2022}. Furthermore, \citet{Anderson2019} showed that CO abundances $\leq10^{-6}$ are needed to explain the observed N$_2$H$^+$ and CO line fluxes of two Upper Sco discs. Secondly, other processes are affecting disc evolution in Upper Sco. Removing the less bound material from the disc outer regions, external photoevaporation halts viscous spreading, making discs smaller and fainter \citep[e.g.][]{Clarke2007}. While in Lupus the irradiation levels are expected to be low \citep{Cleeves2016} and photoevaporation to be negligible (with the possible exception of large discs, \citealt{Haworth2017}), \citet{Trapman2020} argued that the level of irradiation in Upper Sco can be $\approx100$ times higher and photoevaporation more efficient. Thirdly, the initial conditions are different. In this case, Upper Sco cannot be regarded as the subsequent evolutionary stage of Lupus and the two regions must be modelled separately (e.g. if $t_{\rm acc}$ is shorter in Upper Sco than in Lupus, its disc fluxes will decrease faster). 

In the MHD-wind case, models and data have different shapes; no fudge factors are needed to explain the brightest sources, but fainter models require some corrections. Even though \citet{Trapman2022} invoked carbon depletion to reconcile MHD-wind models and the observationally inferred disc sizes in Upper Sco, we note that some of the brightest Upper Sco discs in our dataset were not included in their work because they were  not well resolved (see black-contour dots in Fig.~\ref{fig:C1}). In any case, considering our previous arguments on carbon depletion, our conclusion that MHD-wind models do not need lower CO column densities to match the brightest sources is puzzling. A possible explanation is that our initial disc size distribution is not suitable; in the MHD-wind models of \citet{Tabone2022a,Tabone2022b}, $R_0$ is constant with time, and thus it must match the observed disc sizes in Lupus and Upper Sco. For Lupus, assuming as initial disc size distribution $\log(R_0/{\rm au}) = \mathcal{N}(1.75,\,0.4)$, whose trailing edge agrees with the observationally inferred $R_{\rm CO,68}$ distribution \citep{Sanchis2021}, MHD-wind models require a fudge factor of $\xi^{\rm wind}_{\rm Lup} = 10^{-2}$, similar to the viscous one, to match the data. This is shown in the right panel of Fig.~\ref{fig:3.2}, where models with a reduced CO column density are plotted with a dotted line. As in the viscous case, these models can also reproduce very well the $\dot{M}_{\rm acc}-M_{\rm disc}$ correlation (by construction) and the CO size-luminosity correlation (see Sect.~\ref{app:long2022}). Instead, the very few constraints on the disc size distribution in Upper Sco are not in contrast with our assumption for the initial disc distribution in Sect.~\ref{sec:results}. A possible explanation would be that Upper Sco discs were born more compact than Lupus discs \citep{Barenfeld2016,Barenfeld2017,Miotello2021} or became smaller due to environmental effects such as photoevaporation \citep{Trapman2020}. Better data are needed to draw robust conclusions.

\section{Conclusions}\label{sec:conclusions}
In this Letter we introduced a new method to study protoplanetary disc evolution based on $^{12}$CO fluxes. Assuming optically thick emission, we built a semi-analytical model to compute disc fluxes; the  results agree well (within a factor of three) with those of more time-expensive thermochemical models (\texttt{DALI}). Then, we simulated families of discs, evolving from the same initial conditions either under the effect of viscosity or MHD winds,  and compared their fluxes with Lupus and Upper Sco data. Using fluxes instead of sizes increases our observational samples by a factor of 1.3 in Lupus and 3.6 in Upper Sco, allowing for a more robust comparisons between models and data. 

In the viscous case, our models were brighter than the data. To match the observations, we introduced \textit{\emph{different}} column density fudge factors ($\xi^{\rm visc}_{\rm Lup}=2.5\times10^{-2}$ and $\xi^{\rm visc}_{\rm Sco}=10^{-3}$) that can be explained by carbon depletion (Lupus), and the effects of thermal winds or different initial conditions (Upper Sco). In the MHD-wind case our models matched the brightest discs in Lupus and Upper Sco, but mass-dependent factors were needed to reproduce the fainter sources. In the case of Lupus, when larger initial disc sizes (compatible with the observed distribution) were prescribed, a constant factor ($\xi^{\rm wind}_{\rm Lup}=10^{-2}$, comparable with $\xi^{\rm visc}_{\rm Lup}$) was needed to reproduce the observed fluxes.

Unfortunately, our interpretation of the data is limited by the uncertainties on the initial conditions and the amount of carbon depletion. Nevertheless, our proof of concept shows the usefulness of CO fluxes to study disc evolution. Measuring fluxes instead of sizes is less time-consuming. Additionally, fluxes could be the only accessible observable in farther SFRs. Thanks to   forthcoming surveys that will target tens of discs at limited resolution and with the potential to inform us about carbon depletion, we will be able to obtain the most knowledge about disc evolution from  this new flux-oriented approach.

\begin{acknowledgements}
We are grateful to the referee for their useful comments. This Letter makes use of the following ALMA data: for Lupus ADS/JAO.ALMA\#2015.1.00222.S (PI J. P. Williams), 2016.1.01239.S (PI S. van Terwisga), 2013.1.00226.S (PI K. \"Oberg), 2013.1.01020.S (PI T. Tsukagoshi) and for Upper Sco ADS/JAO.ALMA\#2011.0.00526.S, 2013.1.00395.S (PI J. Carpenter), 2012.1.00743.S (PI G. van der Plas). ALMA is a partnership of ESO (representing its member states), NSF (USA), and NINS (Japan), together with NRC (Canada),  NSC and ASIAA (Taiwan), and KASI (Republic of Korea), in cooperation with the Republic of Chile. The Joint ALMA Observatory is operated by ESO, AUI/NRAO, and NAOJ. We are grateful to L. Trapman for sharing the outputs of his \texttt{DALI} simulations and G. Rosotti for insightful discussions. F.Z. acknowledges support from STFC and Cambridge Trust for a Ph.D. studentship and is grateful to ESO for hosting him for the 2019 Summer Research Programme \citep{messenger_programme} and the 2022 Visitor Programme, when most of this work came to life. 
S.F. is funded by the European Union under the European Union’s Horizon Europe Research \& Innovation Programme 101076613 (UNVEIL). A.M. acknowledges the Deutsche Forschungsgemeinschaft (DFG, German Research Foundation) – Ref no. FOR 2634/2, ER685/11-1. C.F.M. is funded by the European Union under the European Union’s Horizon Europe Research \& Innovation Programme 101039452 (WANDA). Views and opinions expressed are however those of the author(s) only and do not necessarily reflect those of the European Union or the European Research Council. Neither the European Union nor the granting authority can be held responsible for them. This project has received funding from the European Union's Horizon 2020 research and innovation programme under the Marie Sklodowska-Curie grant agreement No 823823 (Dustbusters RISE project). Software: \texttt{numpy} \citep{numpy}, \texttt{matplotlib} \citep{matplotlib}, \texttt{scipy} \citep{scipy}, \texttt{JupyterNotebook} \citep{jupyter_notebook}, \texttt{lifelines} \citep{lifelines}.
\end{acknowledgements}

\bibliographystyle{aa} 
\bibliography{aanda} 

\begin{appendix}\label{app}

\section{Model derivation}\label{app:model}
\begin{figure}[b!]
    \centering
    \includegraphics[width=\columnwidth]{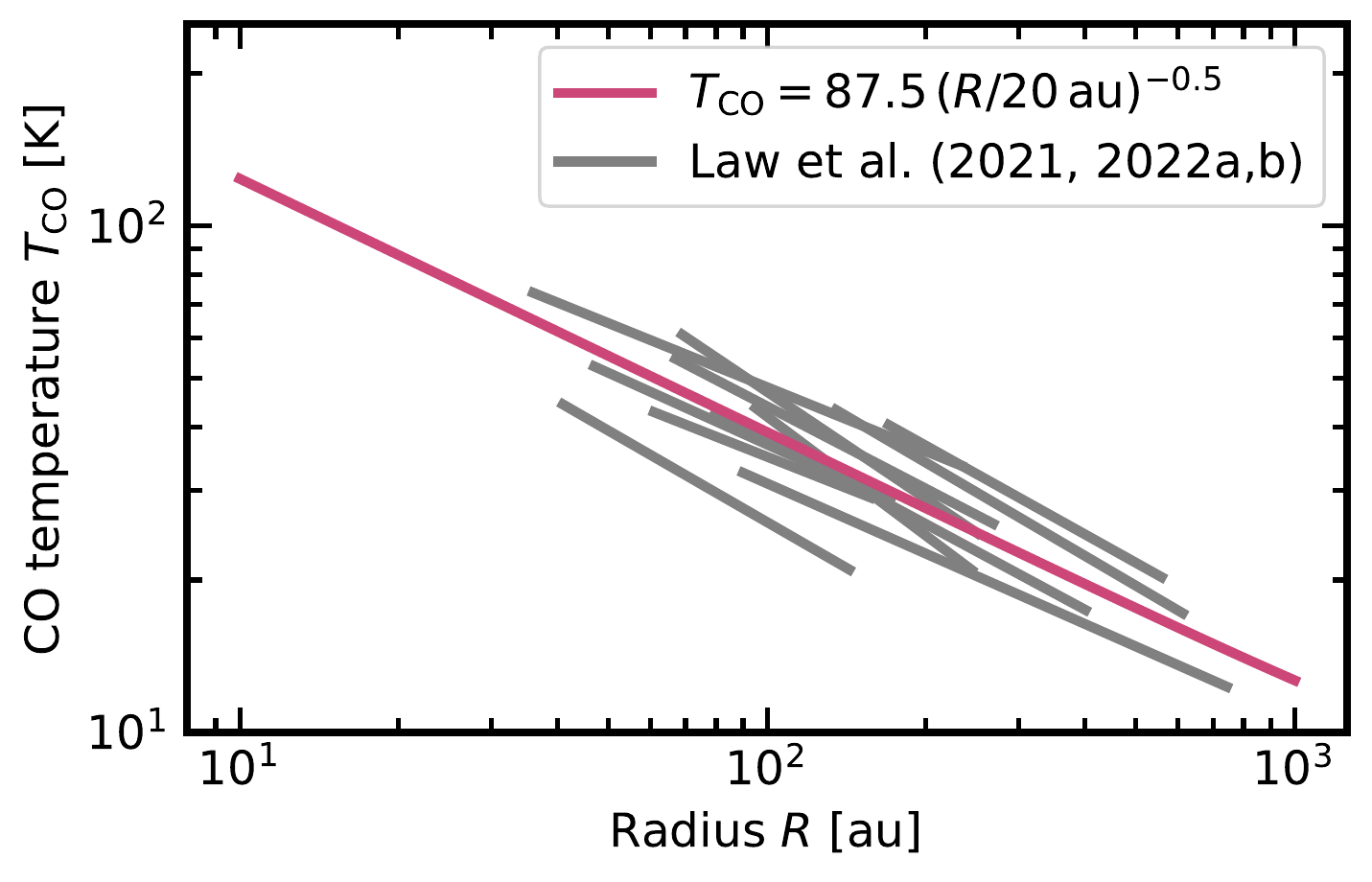}
    \caption{Comparison between Eq.~\ref{eq:A2} (purple line) and the observationally inferred CO temperature profiles (grey). Data: IM~Lup, AS~209, and GM~Aur by \citet{Law2021}; MY~Lup, GW~Lup, WaOph~6, DoAr~25, Sz~91, and CI~Tau by \citet{Law2022}; DM~Tau and LkCa~15 by \citet{Law2023}.}
    \label{fig:A1}
\end{figure}

Under the assumptions that the $^{12}$CO emission is optically thick and in local thermodynamical equilibrium (LTE), the CO brightness at the emitting frequency $\nu_0$ can be computed as
\begin{equation}\label{eq:A1}
    F_{\nu_0} = \dfrac{\cos i}{d^2}\int_{R_{\rm in}}^{R_{\rm out}}B_{\nu_0}(T)\,2\pi RdR,
\end{equation}
where $B_{\nu_0}$ is the black-body emission at temperature $T$,   $R$   is the disc cylindrical radius, $i$ is its inclination, and $d$ is the distance from the observer. We adopted a temperature profile similar to those inferred from CO high-resolution and sensitivity data in T~Tauri discs \citep{Law2021,Law2022,Law2023},
\begin{equation}\label{eq:A2}
    T_{\rm CO} = 87.50\left(\dfrac{R}{20\, {\rm au}}\right)^{-0.5}\, {\rm K},
\end{equation}
and added a temperature floor of $T_{\rm floor}=7\,{\rm K}$, the typical interstellar radiation field in low-mass SFRs:
\begin{equation}
    T^4=T_{\rm CO}^4+T_{\rm floor}^4.
\end{equation}
A comparison of our temperature profile in Eq.~\ref{eq:A2} and those of \citet{Law2021,Law2022,Law2023} is shown in Fig.~\ref{fig:A1}.

We took into account thermal broadening of the $^{12}$CO line as explained by \citet{Rybicki&Lightman1986}. Calling $V$ the gas velocity component along the line of sight, the probability of a CO molecule to be in the velocity range between $V$ and $V + dV$ follows the Maxwell--Boltzmann distribution
\begin{equation}
    p_VdV \propto \exp\left\{-\dfrac{m_{\rm CO}V^2}{2k_{\rm B}T}\right\}dV,
\end{equation}
where $m_{\rm CO}$ is the $^{12}$CO molecular mass, $c$ the speed of light, and $k_{\rm B}$ the Boltzmann constant. The change in frequency (Doppler shift) associated with the velocity $V$ is 
\begin{equation}
    \nu = \nu_0\left(1+\dfrac{V}{c}\right),
\end{equation}
and hence, because $p_\nu d\nu=p_VdV$,
\begin{equation}
     p_\nu d\nu=\dfrac{dV}{d\nu}p_V\left[\dfrac{c(\nu-\nu_0)}{\nu_0}\right]d\nu=\dfrac{c}{\nu_0}p_V\left[\dfrac{c(\nu-\nu_0)}{\nu_0}\right]d\nu.
\end{equation}
Combining this expression with Eq.~\ref{eq:A1} and integrating over the velocity space gives
\begin{equation}
    F_{\rm CO} = \dfrac{\cos i}{d^2}\int_{R_{\rm in}}^{R_{\rm out}}\int_0^\infty I_\nu 2\pi RdRd\nu,
\end{equation}
with
\begin{equation}
    I_\nu d\nu = B_{\nu_0}(T)\exp\left\{-\dfrac{m_{\rm CO}c^2(\nu-\nu_0)^2}{2k_{\rm B}T\nu_0^2}\right\}\dfrac{c}{\nu_0}d\nu,
\end{equation}
which implicitly assumes that optical depth affects the line profile only at the peak (i.e. that the line is not optically thick on a wide velocity range), otherwise its profile would be saturated. The inclination was fixed to the sky-averaged value of $\cos i=\pi/4$, and we adopted $R_{\rm in}=10^{-2}\, {\rm au}$ and $R_{\rm out} = R_{\rm CO}$, the photodissociation radius (see Sect.~\ref{sec:methods}).

\section{Comparison with \texttt{DALI} models}\label{app:comparison}
We benchmarked the results of our semi-analytical model against the disc sizes and fluxes from the thermochemical radiative transfer code \texttt{DALI} published by \citet{Trapman2020,Trapman2022} and \citet{Miotello2016} in the viscous and MHD-wind case, and for different disc inclinations, respectively. We considered a standard diffuse ISM carbon abundance, $x_{\rm CO}=10^{-4}$, and a photodissociation threshold of $N_{\rm CO}=5\times10^{15}\,{\rm cm}^{-2}$. Even though, this value is higher than the standard photodissociation column density of \citet{vanDishoeck&Black1988}, it gives a better agreement between disc sizes in \texttt{DALI} and our model. We tentatively attribute this difference to the effects of freeze-out on the CO column density in \texttt{DALI}.

\begin{figure*}
    \centering
    \includegraphics[width=0.95\textwidth]{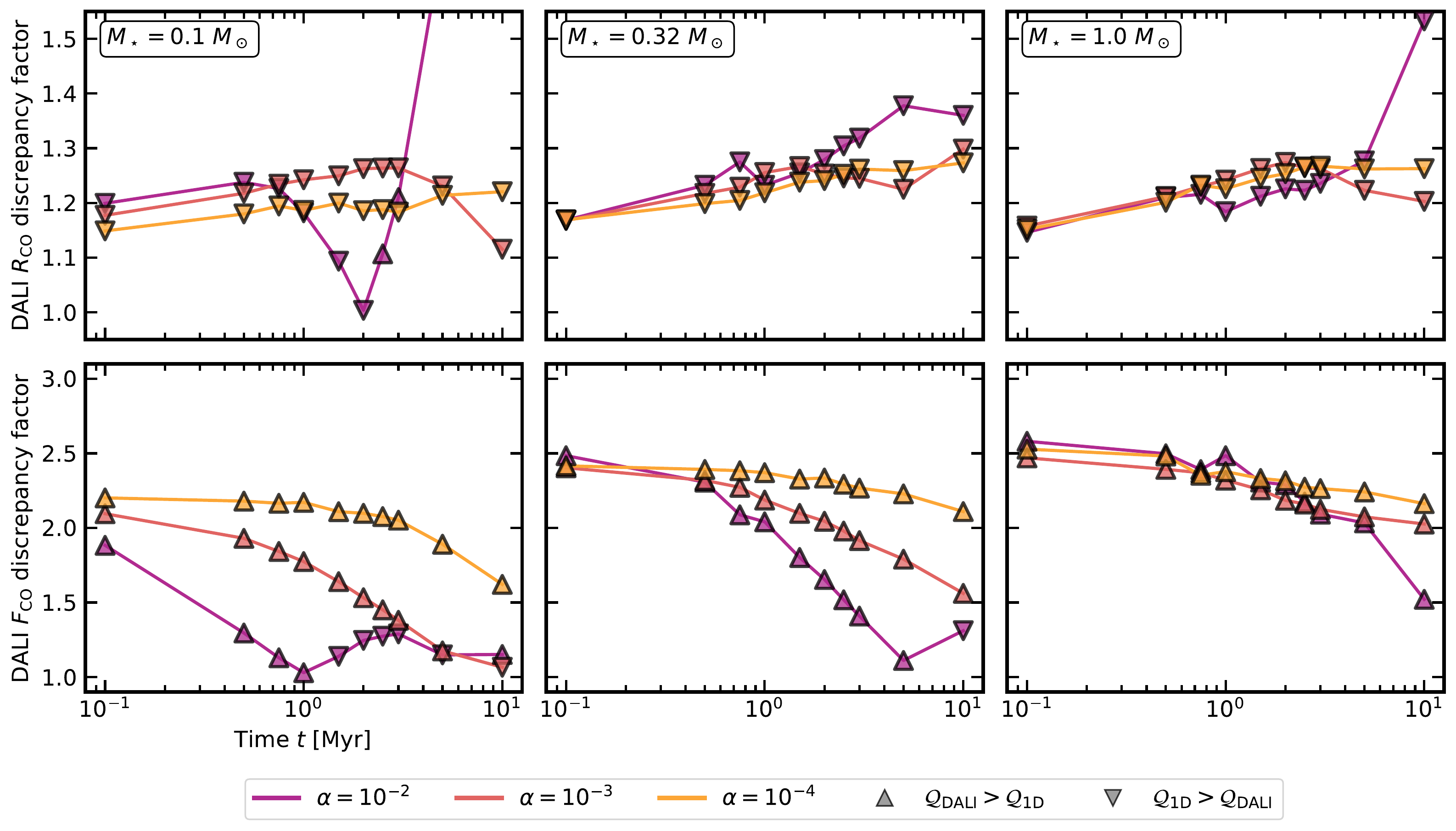}
    \caption{Discrepancy factor (Eq.~\ref{eq:B1}) between the $^{12}$CO $J=2-1$ sizes (upper panels) and fluxes (lower panels) from our semi-analytical model and \texttt{DALI} (from \citealt{Trapman2020}) as a function of time, for a different disc mass and viscous timescale. Upward and downward triangles are used when \texttt{DALI} fluxes overestimate or underestimate our model fluxes, respectively.}
    \label{fig:B1}
\end{figure*}

In the viscous case, sizes and fluxes are from the models of \citet{Trapman2020} for different initial disc masses and viscous timescales (see Table 1 therein), $R_0=10\,{\rm au}$, $d=150\,{\rm pc}$, and $\cos i=1$. Previous to comparison we converted the 90\% CO sizes of \citet{Trapman2020} to $R_{\rm CO}$ using Eq.~F.7 in \citet{Trapman2019} and a power-law temperature profile of $40\,{\rm K}$ at $20\,{\rm au}$ and $-0.25$ exponent. For a given quantity $\mathcal{Q}\in\{R_{\rm CO},F_{\rm CO}\}$ we computed the discrepancy factor between thermochemical model ($\mathcal{Q}_{\rm DALI}$) and our method ($\mathcal{Q}_{\rm 1D}$) results as
\begin{equation}\label{eq:B1}
    \begin{dcases}
        \mathcal{Q}_{\rm DALI}/\mathcal{Q}_{\rm 1D} & {\rm if}\ \mathcal{Q}_{\rm DALI}\geq \mathcal{Q}_{\rm 1D},\\
        \mathcal{Q}_{\rm 1D}/\mathcal{Q}_{\rm DALI} & {\rm otherwise}
    \end{dcases}.
\end{equation}
Our discrepancy factors are shown in the upper and lower panels of Fig.~\ref{fig:B1} for sizes and fluxes. We used a colour gradient for different disc viscosities and the upward or downward triangle when \texttt{DALI} results overestimate or underestimate ours (Eq.~\ref{eq:B1}). Clearly, our models \textit{\emph{overestimate}} \texttt{DALI} disc sizes by less than a factor 1.3 in most cases and \textit{\emph{underestimate}} \texttt{DALI} fluxes by less than a factor 2.5. This difference is due to differences in the line profile because of the high optical depth.

\begin{figure*}
    \centering
    \includegraphics[width=0.95\textwidth]{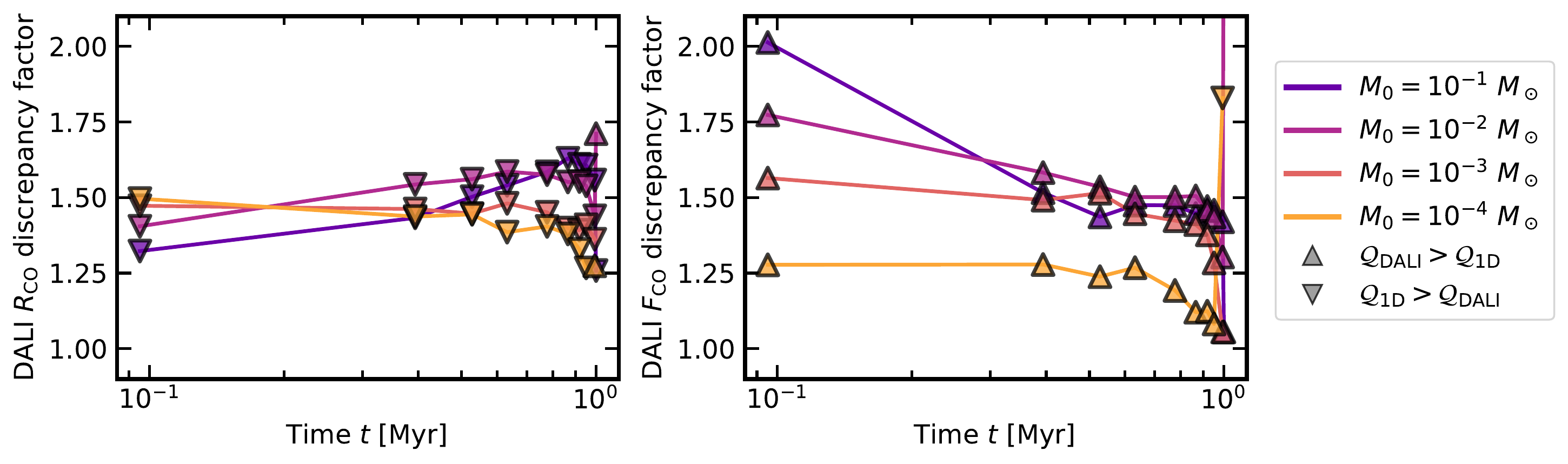}
    \caption{Discrepancy factor (Eq.~\ref{eq:B1}) between the $^{12}$CO $J=2-1$ sizes (right panels) and fluxes (left panels) from our semi-analytical model and \texttt{DALI} (from \citealt{Trapman2022}) as a function of time, for a different disc mass. Upward and downward triangles are used when \texttt{DALI} fluxes overestimate or underestimate our model fluxes, respectively.}
    \label{fig:B2}
\end{figure*}

In the MHD-wind case, sizes and fluxes are from \citet{Trapman2022} for different initial disc masses, $t_{\rm acc}=0.5\, {\rm Myr}$, $R_0=65\, {\rm au}$, $\lambda=3$, $d=150\, {\rm pc}$, and $\cos i=1$. Our discrepancy factors are shown in Fig.~\ref{fig:B2}. Sizes agree well within a factor of 1.75 and fluxes within a factor of two in most cases.

\begin{figure*}
    \centering
    \includegraphics[width=0.95\textwidth]{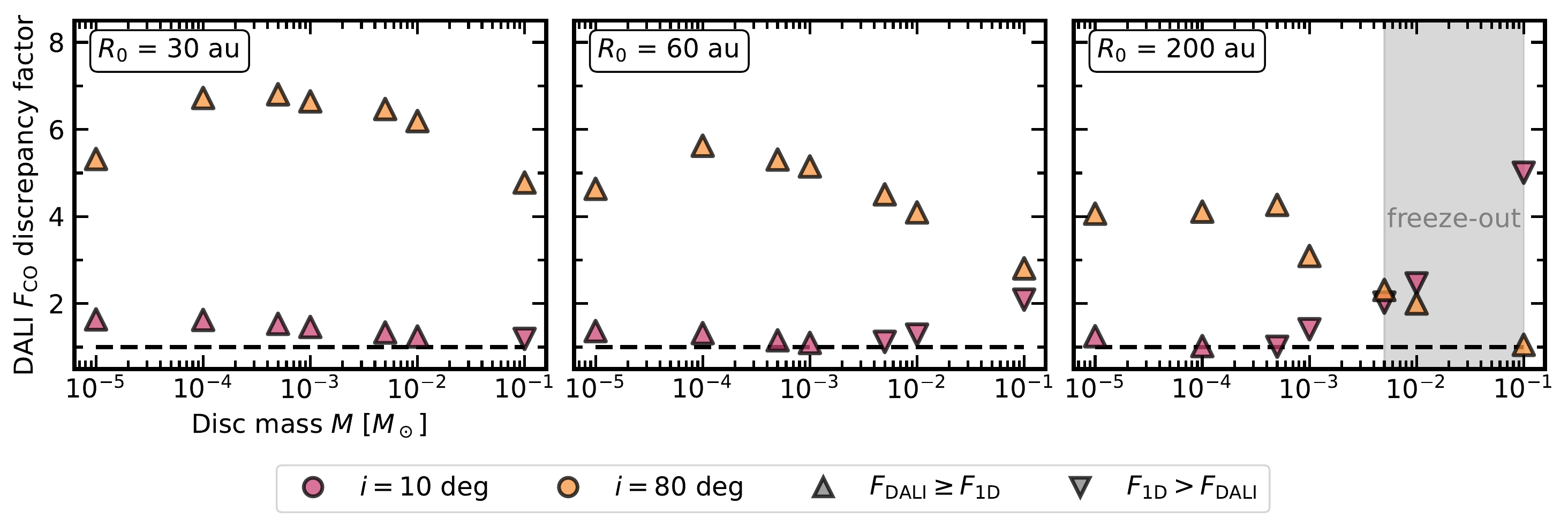}
    \caption{Discrepancy factor (Eq.~\ref{eq:B1}) between the $^{12}$CO $J=2-1$ fluxes from our semi-analytical model and \texttt{DALI} (from \citealt{Miotello2016}) as a function of the disc mass, for a different scale radius and disc inclination. Upward and downward triangles are used when \texttt{DALI} fluxes overestimate or underestimate our model fluxes, respectively.}
    \label{fig:B3}
\end{figure*}

As a final test, we compared our model fluxes with those of \citet{Miotello2016} for different disc inclinations. In this case we adopted a truncated power-law density profile with decay exponent $\gamma=1.5$ \citep{Lynden-Bell&Pringle1974}, on a log-spaced grid with $1\leq\log(R/{\rm au})\leq 4$ and a distance of $100\,{\rm pc}$ \citep{Miotello2016}. Our results are shown in Fig.~\ref{fig:B3}, where the discrepancy factor is plotted as a function of the disc mass for a different scale radius for the $^{12}$CO $J=2-1$ transition at $230.538\,{\rm GHz}$. We used purple and yellow symbols for different disc inclinations (10 and 80 degrees, respectively) and the upward or downward triangle when \texttt{DALI} fluxes overestimate or underestimate our model fluxes (Eq.~\ref{eq:B1}). 

\begin{figure*}
    \centering
    \includegraphics[width=0.95\textwidth]{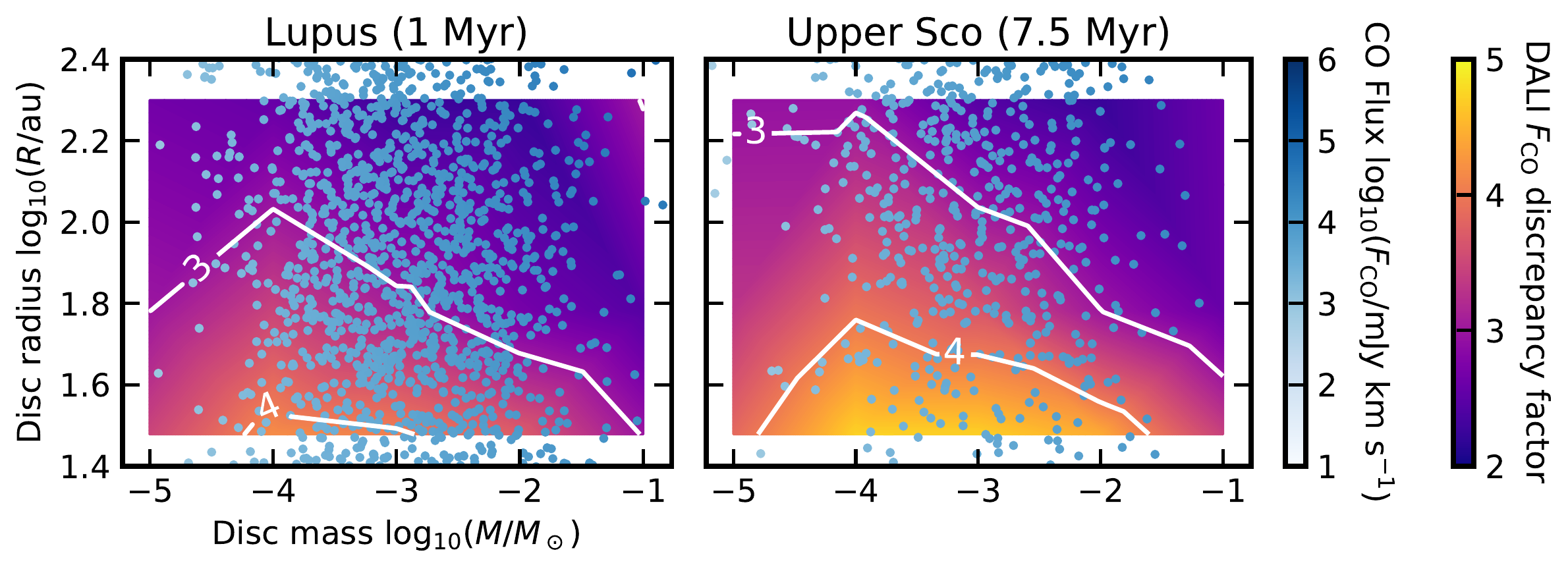}
    \caption{Comparison between the background average discrepancy factor between 1D models and the \texttt{DALI} fluxes of \citet{Miotello2016} (Eq.~\ref{eq:B1}) and the disc population synthesis models, colour-coded by their CO flux, in the same disc mass and scale radius range. Most models are expected to reproduce \texttt{DALI} fluxes within a factor of three.}
    \label{fig:B4}
\end{figure*}

For models close to face-on ($i=10^\circ$), we recover a good agreement between the 1D model and \texttt{DALI} fluxes, with a discrepancy factor of less than two. Instead, for models close to edge-on ($i=80^\circ$), \texttt{DALI} fluxes are larger than ours by a factor of four to six. This can be explained by the increased optical depth through the line of sight, which makes the (otherwise optically thin) outer disc regions more opaque, increasing \texttt{DALI} model fluxes. In Fig.~\ref{fig:B3}, very large and massive discs ($R_0=200\, {\rm au}$, $M_0\geq5\times10^{-3}\, M_\odot$) have a different behaviour, that can be explained by the effects of freeze-out, included in \texttt{DALI} but not in our models. The more massive a disc is, the less efficiently the stellar radiation can penetrate its atmosphere and heat its mid-plane \citep{Miotello2016}; this can cause high-levels of CO freeze-out that make the disc fainter. Larger discs are more prone to freeze-out because more mass resides in the colder outer regions. Even though the optical depth and temperature effects can be taken into account parametrically \citep[e.g.][for freeze-out]{Toci2023}, we decided to keep our model as simple as possible. This is motivated by Fig.~\ref{fig:B4}, where some of our viscous models from Sect.~\ref{sec:results} are plotted over the average \texttt{DALI} discrepancy factor. Most of our models fall in a region of the parameter space where the discrepancy factor is about three. 

We obtained very similar results in the case of the $^{12}$CO $J=3-2$ transition at $345.796\,{\rm GHz}$.

\section{Sample description}\label{app:sample}
In this section we briefly introduce the Lupus and Upper Sco samples taken into account in Sect.~\ref{sec:results}. 

\begin{figure}
    \centering
    \includegraphics[width=0.95\columnwidth]{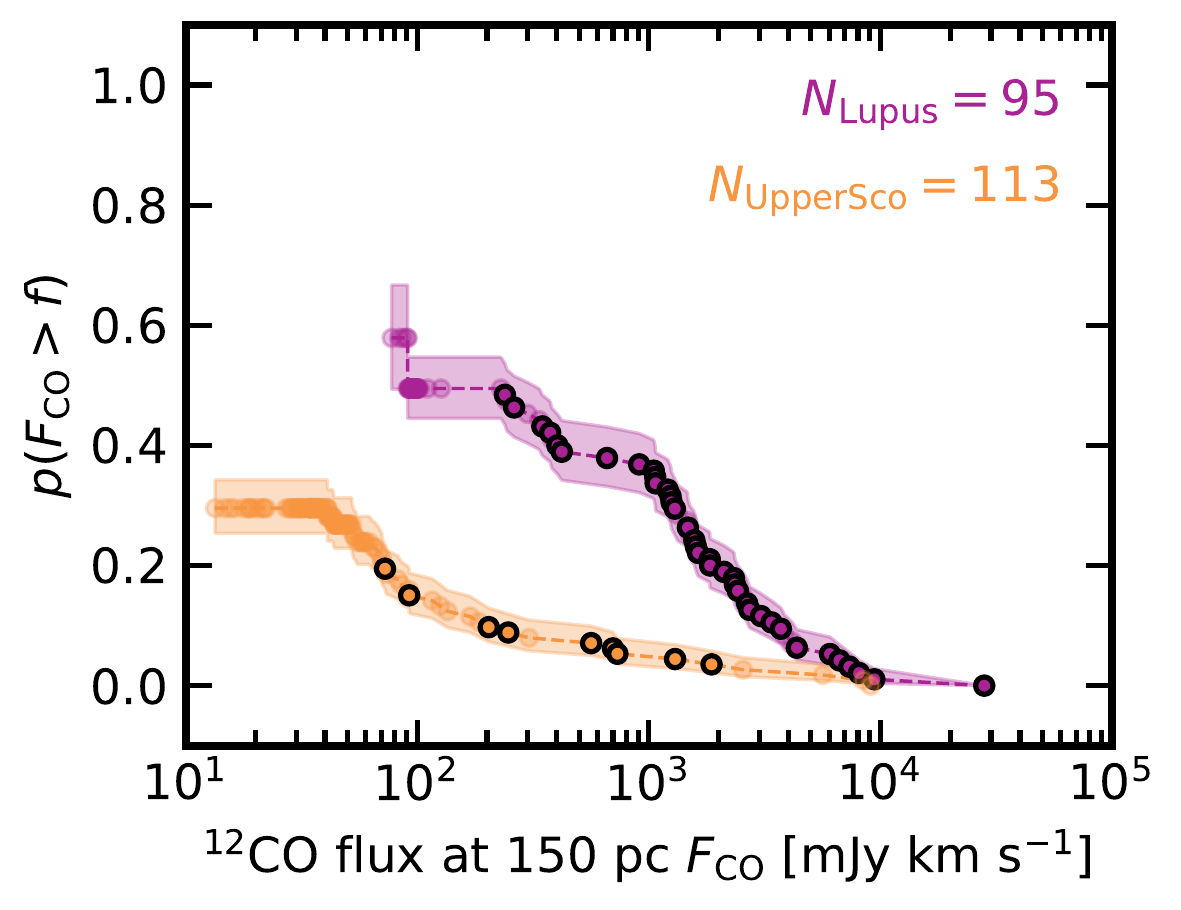}
    \caption{Survival function for Lupus (purple) and Upper Sco (orange). The number of discs targeted by ALMA in each SFR is shown in the same colour in the upper right corner. Resolved discs are plotted with a black contour.}
    \label{fig:C1}
\end{figure}

Lupus discs were observed with ALMA in different programs (\citealt{Ansdell2018,vanTerwisga2018,Cleeves2016,Canovas2016,Sanchis2020}, see summary in Table 1 of \citealt{Sanchis2021}) targeting a total of 100 discs. We have no information on CO fluxes for the five brown-dwarf discs observed by \citet{Sanchis2020}. Of the remaining 95 discs, 48 were detected \citep[>$3\sigma$,][]{Ansdell2018} and 36 resolved \citep{Sanchis2021} in $^{12}$CO. Upper Sco discs were observed with ALMA in different programs \citep{Barenfeld2016,vanderPlas2016} targeting a total of 113 discs. Of these, 32 were detected \citep[>$3\sigma$,][]{Barenfeld2016,vanderPlas2016} and 9 resolved (with well-constrained sizes, \citealt{Barenfeld2017,Trapman2020}) in $^{12}$CO. It is then already clear that considering fluxes instead of sizes increases the sample by a factor of $1.3$ in Lupus and by a remarkable $3.6$ in Upper Sco. We further note that  in Lupus the unresolved discs are all among the faintest sources, while in Upper Sco there are some unresolved discs that are brighter (and potentially larger) than the largest resolved ones. The Lupus surveys targeted the $^{12}$CO $J=2-1$ transition, and the Upper Sco surveys observed the $J=3-2$ transition. To get a homogeneous sample, we rescaled the CO fluxes to the $J=2-1$ rest frequency multiplying by the square of the ratio of the $J=2-1$ to $J=3-2$ frequencies, assuming that the Rayleigh-Jeans approximation holds. We checked this assumption 
for our models and it works well with marginal discrepancies for the largest discs, where the temperatures can be low in the outer regions. We also rescaled the fluxes to a common distance of $150\, {\rm pc}$ using the \textit{Gaia} EDR3 distances  provided by \citet{Manara_ppvii}.

To compare models and observations we made use of the  survival function. For a real-valued random variable $T$, known as lifetime, with probability density function $f$ and cumulative distribution function $F$, the survival function $S$ is defined as
\begin{equation}
    S(t) = p(T > t) = \int_{t}^\infty f(u) du = 1 - F(t).
\end{equation}
For the observational samples, the survival functions were computed considering the CO flux upper limits (left-censored dataset)   using the Kaplan-Mayer estimator built in the Python package \texttt{lifelines} \citep{lifelines} and are shown in Fig.~\ref{fig:C1} in purple and orange for Lupus and Upper Sco. Resolved discs \citep{Barenfeld2017,Sanchis2021} are plotted with a black contour.

We would like to highlight two notes of caution. Firstly, while the Lupus sample is complete (i.e. all young stars with Class II or flat IR excess were observed with ALMA), Upper Sco is not \citep{Luhman&Esplin2020}, which makes the survival function normalisation and the comparison between models and data (see Sect.~\ref{sec:results}) more uncertain. Future surveys observing a larger fraction of Upper Sco stars with discs will make this comparison more reliable. Secondly, a non-negligible fraction of Lupus discs ($\geq17$, splitting equally between detections, 10, and non-detections, 7) are affected by foreground absorption. Instead, \citet{Barenfeld2016} do not report any information on foreground absorption in Upper Sco. For this reason we decided not to take it into account in our analysis.

\section{Comparison with the size--luminosity correlation}\label{app:long2022}

\begin{figure}
    \centering
    \includegraphics[width=\columnwidth]{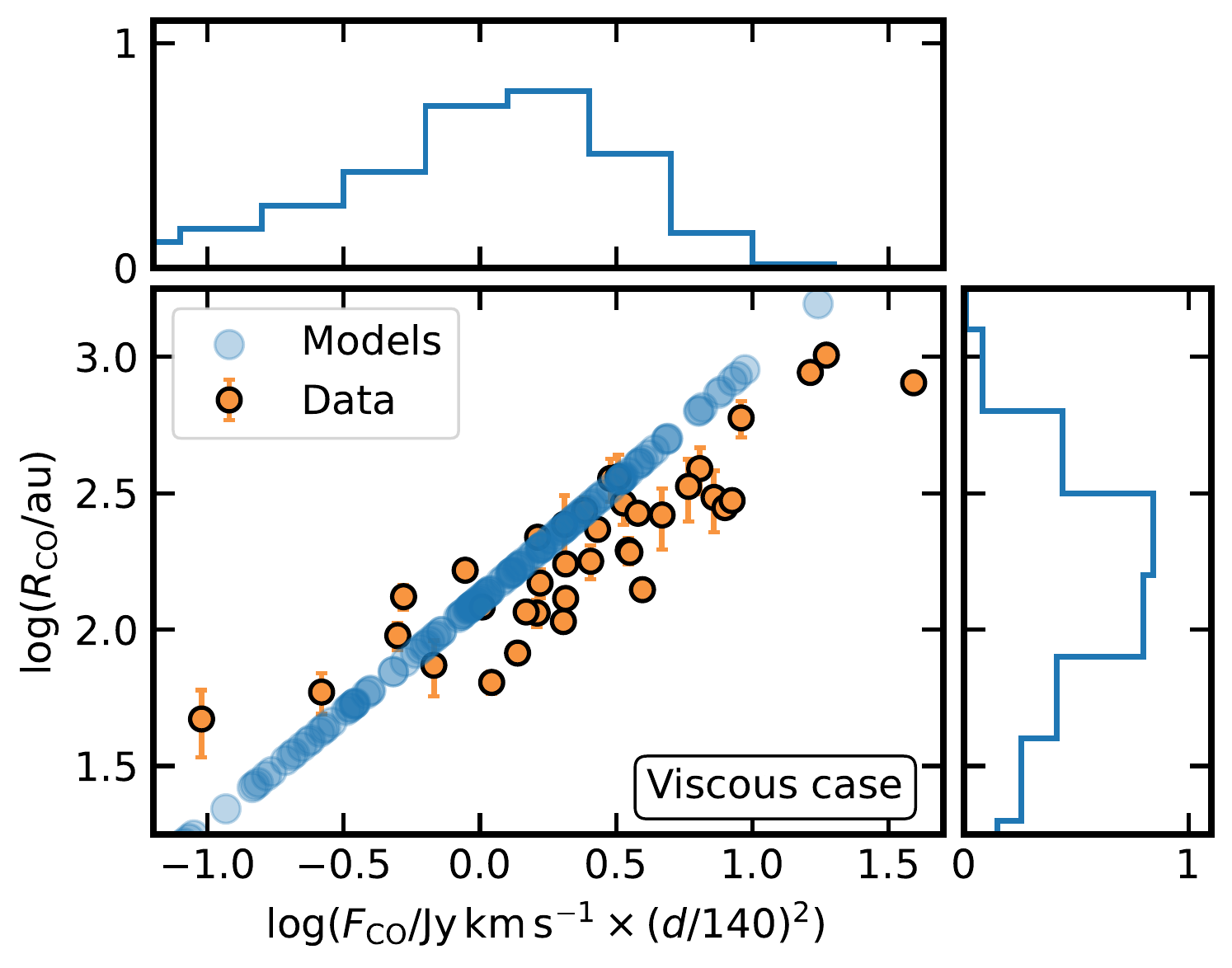}
    \includegraphics[width=\columnwidth]{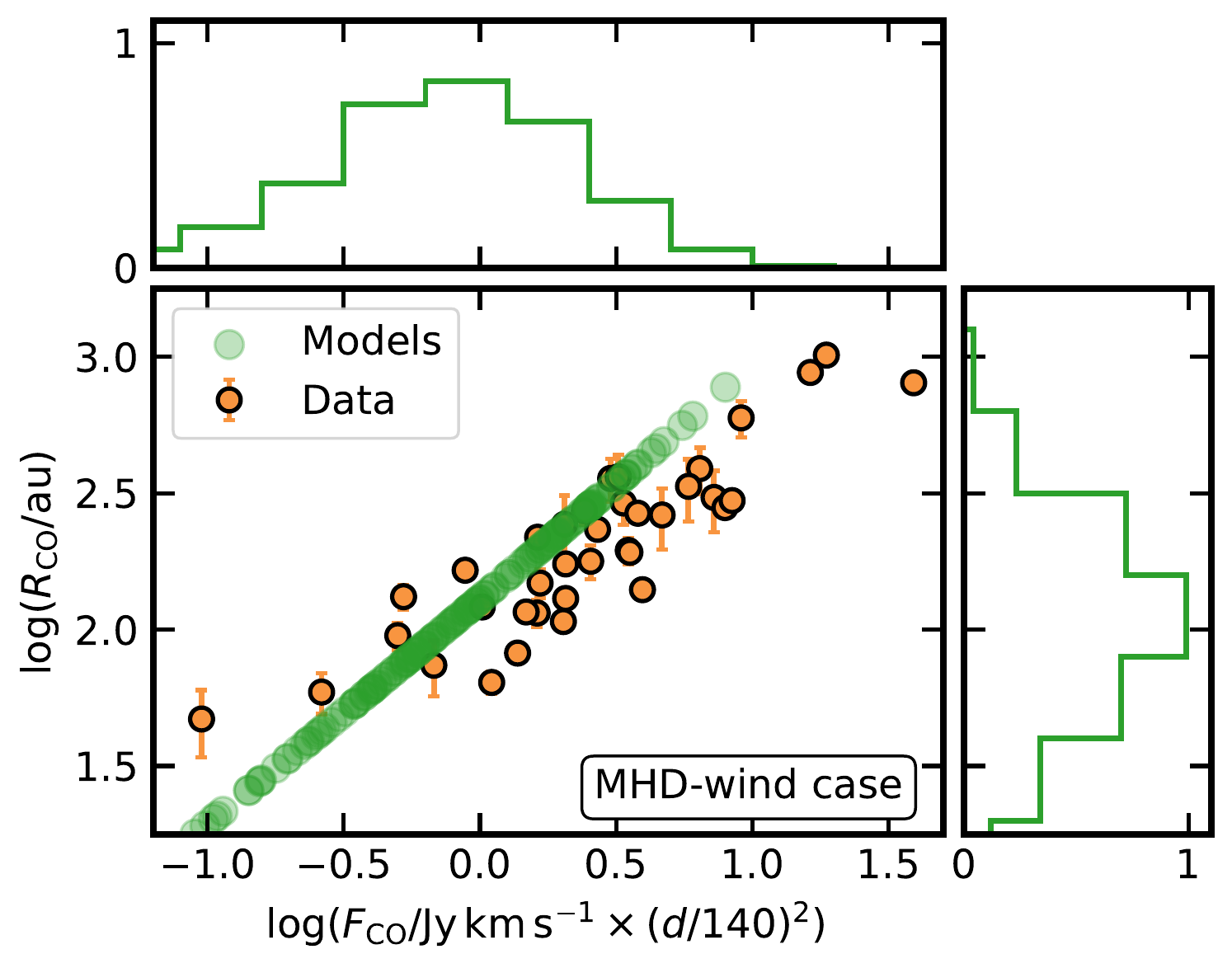}
    \caption{CO size--luminosity correlation. Data are plotted as orange dots and models as blue (viscous case, upper panel) and green (MHD-wind case, bottom panel) dots. Both models can reproduce the correlation slope and roughly its normalisation.}
    \label{fig:D1}
\end{figure}

\citet{Sanchis2021} and \citet{Long2022} showed that for the few sources with well-resolved $^{12}$CO emission, fluxes and sizes are correlated. In Fig.~\ref{fig:D1} our best fit viscous and MHD-wind models from Sect.~\ref{sec:results} (blue and green dots) are plotted in the size--luminosity plane in comparison with \citet{Long2022} data (orange dots), excluding TW~Hya and the Herbig discs. Our models reproduce well the correlation slope and normalisation: they are roughly 1.5 times fainter than the bulk of the data, consistent with the systematic underestimation of our fluxes by a factor of two to three when compared to \texttt{DALI} models. The scatter about the correlation, instead, is underestimated. A better agreement could be obtained introducing a dispersion, for example  in the CO temperature (as in Fig.~\ref{fig:A1}), which is expected to depend on the stellar luminosity and potentially the disc age.

\end{appendix}

\end{document}